\documentclass[aps,prl,nofootinbib,twocolumn,floatfix,nobalancelastpage,
letterpaper,superscriptaddress,showpacs]{revtex4}
\usepackage{graphicx}\usepackage{natbib}\usepackage{amsmath}

\newcommand{\begit}{\begin{itemize}}
\newcommand{\enit}{\end{itemize}}
\newcommand{\begen}{\begin{enumerate}}
\newcommand{\enen}{\end{enumerate}}

\newcommand{\beq}{\begin{equation}}
\newcommand{\eeq}{\end{equation}}
\newcommand{\beqa}{\begin{eqnarray}} 
\newcommand{\eeqa}{\end{eqnarray}} 

\begin{document}

\title{Upper Bound on the Energy of Particles and Their Secondary Neutrinos}

\author{Todd A.~Thompson}
\author{Brian C.~Lacki}
\affiliation{Department of Astronomy, Ohio State University, Columbus, Ohio 43210}
\affiliation{Center for Cosmology and Astro-Particle Physics, Ohio State University, Columbus, Ohio 43210}

\date{March 9, 2011}

\begin{abstract} 
We derive an upper limit to the energy of nuclei accelerated via the Fermi mechanism in any relativistic shockwave, driven by any astrophysical engine. This bound is accessible to current and upcoming ultra-high energy neutrino experiments. Detection of a single neutrino with energy above the upper limit would exclude all sites of shock acceleration, and imply physics beyond the Standard Model. We comment on the possibility that relativistic flows launched by supermassive black hole mergers are the source of the observed ultra-high energy cosmic rays. 
\end{abstract}

\pacs{95.85.Ry, 98.70.Rz, 98.70.-f}

\maketitle

The origin of ultra-high energy cosmic rays ($>10^{18.5}$\,eV; UHECRs) is unknown.  
Broadly speaking, they can be produced by 
shock acceleration via the Fermi mechanism
\cite{fermi,hillas,waxman95,vietri95}, 
direct acceleration by the electromagnetic fields
of compact objects 
\cite{gunn_ostriker,lovelace,kardashev,arons},
and the decays of hypothetical super-massive particles
beyond the Standard Model (``top-down'' scenarios) 
\cite{bhatt,protheroe_stanev}.
Fermi acceleration is thought to operate in supernova 
remnants, producing the
low energy CRs ($\lesssim10^{15.5}$\,eV) in the Galaxy.  
Similarly, the relativistic shocks of gamma-ray 
bursts and the flares of active galactic nuclei 
are leading prospective astrophysical 
engines for UHECR acceleration \cite{waxman95,lovelace}.

Whatever their source, 
UHECR acceleration must occur in the local universe.  
If UHECRs are protons, they are attenuated on a scale 
of $\lesssim$\,100 Mpc for energies $\gtrsim5\times10^{19}$\,eV 
by the  Greisen-Zatsepin-Kuz'min (GZK; \cite{gzk})
mechanism, producing secondary particles that
decay to  $e^{\pm}$, $\gamma$-rays, 
and UHE neutrinos. If UHECRs are nuclei,
photodisintegration on the cosmic infrared background 
suppresses the flux at a similar energy 
\cite{stecker}.

Although a decrease in the UHECR flux above
the GZK threshold has now been observed \cite{hires,auger:gzk}, 
it is unknown whether this is evidence of
UHECR attenuation,
or instead an intrinsic limitation to 
the acceleration process in astrophysical engines.
If the former, then secondary UHE neutrinos 
with energies above the GZK cutoff should propagate to 
us unimpeded from across the universe \cite{weiler,seckel}.  
Since experimental searches for UHE neutrinos 
are sensitive up to $\sim10^{26}$\,eV \cite{forte},
it is of considerable interest to understand if physical
bounds can be placed on the acceleration process that 
limit the maximum energy of the primary UHECR spectrum.

Such a bound has been constructed for the case of 
direct electromagnetic acceleration \cite{kardashev},
but no such bound has been formulated
for the case of relativistic shock acceleration.
Here, we show that the 
 physics of the Fermi mechanism and synchrotron cooling,
together with an absolute upper bound 
on the bolometric luminosity of any astrophysical engine,
combine to produce an experimentally testable
upper bound on the energy of the UHECR population in the universe.

\textit{The Maximum Energy of Particles.}---
We consider particle acceleration in a 
flow of total luminosity $L$ and 
bulk Lorentz factor $\Gamma$.   As in Ref.\ \cite{waxman95}, 
we assume that the central engine is transient and stationary
in the observer frame, and  that 
the region of particle acceleration is a shell of radial extent 
$\delta r=\delta r^\prime/\Gamma= r/\Gamma^2$, and perpendicular extent
$r_\perp^\prime=r_\perp$, where $r$ is the 
distance from the shell to the engine, and where primes denote
quantities in the flow rest frame.  The connection
between $\delta r$, $r$, and $\Gamma$ follows from the
fact that an unsteady flow of thin shell geometry 
traveling at $c-\delta v$, with
$\delta v\ll c$, broadens kinematically
to a thickness $\delta r\approx r \delta v/c$.

The comoving magnetic energy density in the 
shell is $u_B^\prime=\epsilon_{\rm B} L/(4\pi \Psi r^2 \Gamma^2 \beta c)$, 
where $\beta=v/c$ is the shock velocity,
$\epsilon_{\rm B}(\leq1)$ measures the efficiency of
magnetic field generation, and $\Psi(\leq1)=\Omega/4\pi$ measures 
the beaming of the flow into solid angle $\Omega$.
A necessary requirement for acceleration is that 
\cite{hillas,waxman95}
$2\pi fR_L^\prime < \min[\delta r^\prime,r_\perp^\prime]$,
where $R^\prime_L=E^\prime/(ZeB^\prime)$ is the Larmor radius,
$E^\prime=E/\Gamma=\gamma^\prime mc^2$ is the particle energy, 
$Ze$ is the charge, $B^\prime$ is the magnetic field strength, 
and $f$ is a constant of order unity. 
We first focus on the case with $\delta r^\prime<r_\perp^\prime$,
and return to the physics of more highly beamed flows below.
Assuming $\beta\approx1$, which we confirm below, and
combining these expressions, one obtains an upper bound
on the accelerated particle energy \cite{waxman95}:
\beq
E_{\rm max}\lesssim\frac{Ze}{f\pi\Gamma}
\left(\frac{\epsilon_{\rm B} L}{2\Psi c}\right)^{1/2}.
\label{emax1}
\eeq

However, there are two crucial limitations to Eq.~(\ref{emax1}).  
First, the same magnetic field that confines the particles to the accelerating
region also causes radiative losses via synchrotron.  Acceleration to $E$
requires that the acceleration time $t^\prime_{\rm acc}\sim R_{\rm L}^\prime/c$ 
is less than the synchrotron cooling time, $t^\prime_{\rm synch}$, 
which sets a lower limit on $\Gamma$ \cite{waxman95}:
\beq
\Gamma>\left(\frac{E}{m_p c^2}\right)^{2/5}
\left(\frac{8\pi f (Ze)^3 }{9 m^2 c^5 \delta t}\right)^{1/5}
\left(\frac{2\epsilon_{\rm B} L}{\Psi c}\right)^{1/10},
\label{gamma1}
\eeq
where $m$ is the particle mass,
$\delta t\sim r/(\Gamma^2 c)$ is the observed variability 
timescale of the system, and we have assumed that 
$v^\prime/c=1$, where $v^\prime$
is the particle velocity in the rest frame 
of the flow. Because $t^\prime_{\rm acc}\propto R_L^\prime$, 
the condition $t^\prime_{\rm acc}<t^\prime_{\rm synch}$ implies that
radiation reaction can be ignored in deriving Eq.~(\ref{gamma1})
\cite{nelson_wasserman}.
 
The second fundamental constraint on Eq.~(\ref{emax1})
follows from the fact that there is a maximum $L$ attainable
by any astrophysical engine \cite{mtw}:
\beq
L_{\rm max}\sim c^5/G\,\simeq\,3.63\times10^{59}\,\,{\rm ergs\,\,s^{-1}},
\label{lmax}
\eeq
equivalent to radiating the entire 
rest mass of a body in a single light travel time
across its Schwarzschild radius.   $L_{\rm max}$ is only 
approached during the merging of  
binary BHs in gravity waves. We return below to 
whether $L_{\rm max}$ in gravity waves 
can couple to electromagnetic fields, thus producing
$u_B^\prime\propto\epsilon_{\rm B} L$ with $\epsilon_{\rm B}>0$.
Here, it is sufficient to note
that $L\lesssim L_{\rm max}$
for any astrophysical engine.

The limits on $\Gamma$ and  $L$ combine with Eq.~(\ref{emax1}) 
to produce a unique $E_{\rm max}$ at a critical $\Gamma_{\rm crit}$. 
We find that 
\beq
E_{\rm max}(\delta t)=
\left[\frac{9}{64\pi^6}\frac{\epsilon_{\rm B}^2}
{f^6}
\frac{m^4 c^{17} Z^2 e^2\,\delta t}{\Psi^2G^2}\right]^{1/7}
\label{emax}
\eeq
is the maximum energy of particles 
accelerated in relativistic shocks driven by
maximally luminous astrophysical engines.  
To evaluate $E_{\rm max}$, note that 
the observed variability timescale $\delta t=t/\Gamma^2$,
where $t$ is the light-travel time between the engine
and the acceleration region.  Since $t$ is bounded by the 
time for GZK losses  for
protons ($t_{\rm loss}$), or photodisintegration losses for nuclei,
\beq
\hspace{-.08cm}E_{\rm max}(t)\approx1\times10^{24}\,{\rm eV}
\left(\frac{A^4\,\epsilon_{\rm B}}{\Psi f^4}
\frac{t}{t_{\rm loss}}\right)^{\hspace{-.1cm}1/5}
\label{emax_hubble}
\eeq
and 
$\Gamma_{\rm crit}\approx250\,Z 
(\epsilon_{\rm B}/f^{2/3}\Psi)^{3/10}
(m_p/m)^{4/5}(t_{\rm loss}/t)^{1/5}$,
where $A=m/m_p$, $t_{\rm loss}\sim50$\,Myr
for protons or nuclei \cite{kelner2,stecker_salamon}.
For $\Gamma\ne\Gamma_{\rm crit}$,
particles are not accelerated to $E_{\rm max}$.
Because the classical synchrotron formula used in 
Eq.~(\ref{gamma1}) is invalidated when the energy of the emitted
photons exceeds the particle energy \cite{landau}, we
verified that $\gamma^\prime B^\prime/[m_p^2 c^3/(\hbar Ze)]\ll1$
for $\Gamma\ge1$ and $t\gtrsim10^{-4}$\,s, 
confirming our use of the classical formula
in deriving Eq.~(\ref{emax_hubble}) for nuclei;
electron cooling will be modified by QED effects,
but we do not consider them here.  

Protons reaching $E_{\rm max}$ lose energy 
via photomeson processes, 
producing secondary neutrinos.  Although the
maximum neutrino energy is $E_{\rm max}$ \cite{kelner1, kelner2},
the average energy is
$E_{\rm max,\,\nu}\sim E_{\rm max}/(20A)$ 
\cite{kelner1,atoyan_dermer,murase_beacom};
that is, 
\beq
E_{\rm max,\,\nu}(t)\sim
5\times10^{22}\,\,{\rm eV}\,
\left(\frac{\epsilon_{\rm B}}{\Psi f^4}
\frac{1}{A}
\frac{t}{t_{\rm loss}}\right)^{1/5}\hspace{-.2cm},
\label{emax_hubble_nu}
\eeq
which is maximized for protons.

Although Eq.~(\ref{emax_hubble}) establishes
an upper bound on $E$, astrophysical engines may not typically
drive outflows with $t=t_{\rm loss}\sim50$\,Myr. Indeed,
unsteady relativistic flows 
develop internal shocks, dissipating their energy 
on a scale $r=\Gamma^2c\delta t$
\cite{meszaros_rees}. To evaluate Eq.\ (\ref{emax}) 
in this limit, we scale the variability
timescale $\delta t$ to the light-crossing time of
the largest compact objects,
super-massive BHs of $\sim10^9$\,M$_\odot$ 
($t_L\sim2 GM/c^3\sim10^4$\,s):
\beq
\hspace*{-.25cm}E_{\rm max}(\delta t)\sim1\times10^{23}\,\,{\rm eV}\,
\hspace*{-.15cm}\left(\frac{A^4Z^2\epsilon_{\rm B}^2}{f^3\Psi^2}
\frac{\delta t}{10^4{\rm s}}\right)^{\hspace*{-.08cm}1/7}
\hspace*{-.25cm},
\label{emax_BH}
\eeq
\beq
E_{\rm max,\,\nu}(\delta t)\sim6\times10^{21}\,\,{\rm eV}
\left(\frac{Z^2\epsilon_{\rm B}^2}{A^3f^3\Psi^2}
\,\frac{\delta t}{10^4{\rm s}}\right)^{1/7}\hspace*{-.15cm},
\label{emax_BH_nu}
\eeq
$\Gamma_{\rm crit}\approx2020\,\,Z^{5/7}
A^{-4/7}(10^4\,{\rm s}/\delta t)^{1/7}(\epsilon_{\rm B}/
\Psi)^{3/14} f^{-1/7}$.
For the $10$\,M$_\odot$ BHs that 
result from the collapse of some massive stars, 
$t_L\sim10^{-4}$\,s, and 
$E_{\rm max}\sim8.6\times10^{21}$\,eV for protons.
Note the dependence on $A$ and $Z$; 
for $Z\propto A$, $E_{\rm max}(\delta t)\propto A^{6/7}$, so
that $E_{\rm max}$ is $\approx30$ times 
higher for iron nuclei than for protons.
 
UHE neutrinos may also come directly from the decay
of unstable particles in the
flow (e.g., $\pi^{\pm}$). 
Requiring $t_{\rm acc}^\prime<\tau\gamma^\prime$,
where $\tau$ is the particle decay time, 
$\tau>8\times10^{-4}\,{\rm s}\,(\delta t/10^{4}\,{\rm s})^{4/7}
(\epsilon_{\rm B}Z^8/A^5f^4\Psi)^{1/7}$
is required for acceleration. 
Thus, $\mu^{\pm}$, $\pi^{\pm}$, and $K^{\pm}$ 
may be directly accelerated, but only for
$\delta t\ll10^4\,{\rm s}$,
implying that it is impossible to exceed the limit of
Eqs.~(\ref{emax_BH}) and (\ref{emax_BH_nu}) by directly 
accelerated unstable particles.

{\it Extreme Beaming.---} 
All of the above expressions are valid 
for the case of a beamed flow with $\delta r^\prime < r_\perp^\prime$.
The opposite limit requires that $\Gamma\gtrsim1/(2\Psi^{1/2})$, and 
for particles to be accelerated one requires $R_L^\prime < r_\perp^\prime$.
These conditions lead to a unique $E_{\rm max}$ and a minimum
Lorentz factor $\Gamma_{\rm min}$ above which this limit obtains:
\beq
E_{\rm max}(t,\Gamma\gtrsim1/2\Psi^{1/2})
\sim Z\alpha^{1/2}E_{\rm pl}\sim Z\,10^{27}\,\,{\rm eV}, 
\label{kardashev}
\eeq
and $\Gamma_{\rm min}\simeq
3\times 10^6 (Z^5/f)^{1/2}\epsilon_{\rm B}^{3/4} A^{-2} (t_{\rm loss}/t)^{1/2}$,
where $E_{\rm pl}$ is the Planck energy and $\alpha$ is the fine structure
constant.  Equation (\ref{kardashev}) is identical to the bound for loss-less 
electromagnetic acceleration \cite{kardashev}.  We have not emphasized this 
case here since $\Gamma_{\rm min}$ is very high, implying 
a solid angle for the flow of $\lesssim10^{-12}$\,sr.

\begin{figure}[t!]
\includegraphics[width=\columnwidth,clip=true]{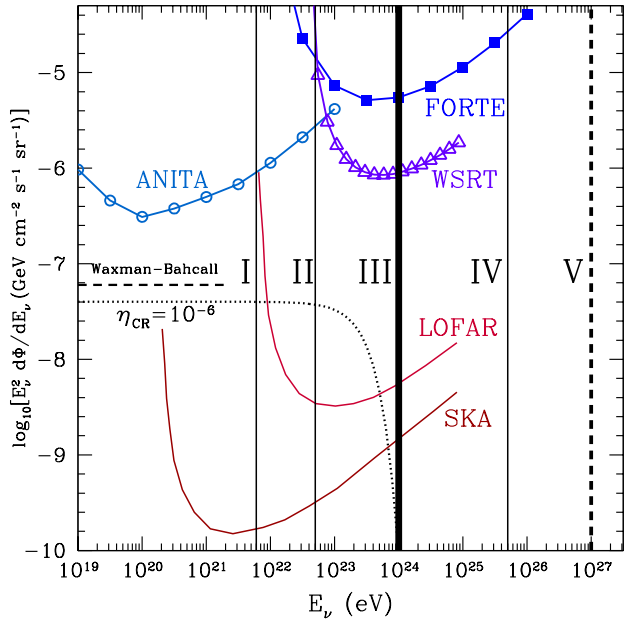}
\caption{Neutrino intensity versus energy.  
Current (ANITA, WSRT, FORTE) and projected (LOFAR, SKA) 
experimental upper bounds are indicated.   
The upper limits on accelerated particle energy derived 
here are indicated by the solid vertical lines 
I-V. Lines III and V (heavy) show the absolute upper 
bounds for spherical and highly-beamed flows 
(Eqs.~\ref{emax_hubble} and \ref{kardashev}), 
beyond which no particle can be accelerated by a 
relativistic shock launched from any astrophysical engine.
Lines II and IV mark the average secondary neutrino energies
expected for these two limiting cases.
Line I corresponds to Eq.~(\ref{emax_BH_nu}).
\label{plot}}
\end{figure}

\textit{Discussion.}---  
Our primary result is that for maximally
efficient acceleration in relativistic shocks, driven by
maximally luminous astrophysical engines, the maximum particle
energy is given by Eq.\ (\ref{emax_hubble}).  For extremely
highly beamed flows  (likely unphysically so), 
the bound is given by Eq.\ (\ref{kardashev}).
These bounds yield
maximum and average secondary neutrino energies that are
within reach of current and upcoming experiments.

Figure \ref{plot} shows experimental upper bounds on the UHE
neutrino intensity above $10^{19}$\,eV from ANITA \cite{anita1}, WSRT\cite{wsrt}, 
and FORTE \cite{forte}, and projected bounds for SKA and LOFAR \cite{scholten}.   
The Waxman-Bahcall flux is shown for reference (dashed line) \cite{waxman_bahcall}. 
The vertical solid lines show the bounds on 
energy derived here assuming $A=Z=f=\epsilon_{\rm B}=1$. 
Lines I and II
show $E_{\rm max,\,\nu}(\delta t=10^{4}\,{\rm s})$ (Eq.~\ref{emax_BH_nu})
and  $E_{\rm max,\,\nu}(t=t_{\rm loss})$ (Eq.~\ref{emax_hubble_nu}).
The heaviest vertical solid line III marks $E_{\rm max}(t=t_{\rm loss})$ 
(Eq.~\ref{emax_hubble}), the absolute upper bound on UHE neutrino energy
for spherical shocks ($\Psi=1$).
For acceleration to line III, the secondary neutrino spectrum 
drops sharply at $E_{\rm max,\,\nu}\sim E_{\rm max}/20$ 
(line II), as does the 
heavy dotted curve  (see below).  The absolute bound for beamed
flows is also shown (line V, heavy dashed; Eq.~\ref{kardashev}), 
as is $E_{\rm max,\,\nu}$
for such particles (line IV).

Figure \ref{plot} shows that the upper bounds presented here are testable.  
If the primary UHECRs that produce the yet-to-be-measured
UHE neutrino flux are in fact accelerated by relativistic
shocks,  we predict a dramatic downturn in the UHE neutrino flux 
at or below $E_{\rm max,\,\nu}\simeq5\times10^{22}\,\,{\rm eV}$ 
(line II; Eq.~\ref{emax_hubble_nu}), 
and most likely well below $\sim10^{22}\,\,{\rm eV}$, since our supposition
is that actual astrophysical engines fail to reach $L_{\rm max}$,
have $\epsilon_{\rm B}\ll1$, and may not have $\Gamma=\Gamma_{\rm crit}$,
or $t=t_{\rm loss}$ (Eq.~\ref{emax_hubble}). 
In turn, the observation of a cutoff in the UHE
neutrino spectrum will put limits on the sources'
overall power, since $L$ is directly connected with $E_{\rm max}$ 
\cite{waxman95}. In particular, we can invert the expressions
above to provide a lower limit on the sources' luminosity, given
the observation of a single UHECR of energy $E_{21}=E/10^{21}\,{\rm eV}$:
\beq
L\gtrsim2\times10^{54}\,{\rm ergs\,\,s^{-1}}\,E_{21}^{7/2}
\frac{\Psi}{ZA^2}\frac{f^3}{\epsilon_{\rm B}}
\left(\frac{1\,{\rm s}}{\delta t}\right)^{\hspace{-0.05cm} 1/2}.
\eeq

Importantly, the observation of a single neutrino above $E_{\rm max}$ 
(line III) immediately rules out spherical shock acceleration as the 
origin of the primary particle's high energy.  Indeed, since even for 
a mono-energetic primary proton spectrum at $E_{\rm max}$ (III), the 
vast majority of the secondary neutrinos will have $E_{\rm max,\,\nu}$ 
(II), the detection of neutrinos above $E_{\rm max,\,\nu}$ strongly 
indicates that the primary was not accelerated by a relativistic shock.
The remaining possibilities are that (1) the flow was exceedingly highly 
beamed ($\Omega<10^{-12}$\,sr), (2) QED effects suppressed synchrotron cooling (requiring
$t<10^{-4}$\,s at $L=L_{\rm max}$), (3) the primary particle was produced by an unknown class of 
near-maximal loss-less electromagnetic accelerators, as in \cite{kardashev}, 
or (4) the detected neutrino was generated by the
decay of a super-massive primary from beyond the 
Standard Model \cite{bhatt,protheroe_stanev}.

\textit{BH-BH Mergers as UHECR Sources?}---
The fact that $L_{\rm max}$ (Eq.~\ref{lmax}) 
is only thought to be approached in 
gravity waves during BH-BH mergers motivates a consideration 
of these events as the primary source of the UHECRs.

If we assume that the galaxy merger rate at $z=0$ 
($\sim0.03$\,Gyr$^{-1}$ per galaxy; \cite{hopkins})
corresponds to the merger rate of 
$M_{\rm BH}\sim10^8$\,M$_\odot$ BHs, that $\eta_{\rm CR} M_{\rm BH}c^2$ 
is radiated in UHECRs per merger, and that the local number density of
large galaxies like  the Milky Way is $10^{-2}$\,Mpc$^{-3}$,
by comparing the volumetric power produced in CRs 
with the inferred UHECR production rate per comoving volume ($\approx0.7\times10^{44}$
ergs yr$^{-1}$ Mpc$^{-3}$; \cite{waxman_bahcall}) one finds  
that $\eta_{\rm CR}\sim 10^{-6}$ is required if such mergers are
to dominate UHECR production.  Assuming an $E^{-2}$ UHECR spectrum 
to $E_{\rm max}(t=t_{\rm loss})$, in Figure 1 we sketch the expected 
neutrino spectrum ({\it dotted curve}). Taking the local GZK
volume to be $\sim(100\,{\rm Mpc})^3$, 
the local merger rate is of order 0.3\,Myr$^{-1}$,
implying that 
$\sim100$ BH-BH merger events would contribute to 
the observed UHECRs.  Although such particles would be expected to 
trace the large-scale distribution of galaxies on the sky,
as has been claimed by Auger \cite{auger_correlation},
the low source density and high energy release per event
may be ruled out if the UHECRs are protons \cite{murase_takami}
(see Ref.\ \cite{auger_composition}). 

Stellar-mass BH-BH mergers might also contribute to the
UHECR budget, but given the inferred rate per galaxy 
($\sim10^{-8}-10^{-6}$\,yr$^{-1}$ \cite{belczynski})
the UHECR energy production rate per volume is smaller than for
supermassive BH-BH mergers by $\sim10^3-10^5$.
Nevertheless,  since these sources are
transient and potentially nearby, 
they could in principle contribute to the 
observed UHECR budget.

Can BH-BH mergers accelerate CRs?
Simulations have argued that the efficiency
of production of electromagnetic waves 
in the vicinity of BH-BH mergers may be large, 
but only for ambient magnetic field strengths far
in excess of those thought to accompany accretion \cite{palenzuela}.
Thus, $\epsilon_{\rm B}$ in our expressions
is the primary unknown in forwarding BH-BH mergers as
the source of UHECRs.

{\it Consequences of Maximal Shocks.---}
If any astrophysical engine, be it merging BHs or otherwise,
reaches $L_{\rm max}$, with $\Gamma=\Gamma_{\rm crit}$,
there are observational consequences beyond the direct detection of
UHECRs or neutrinos.  In particular, the particles 
will produce synchrotron radiation 
with observed photon energy $E_{\rm synch}
=\Gamma h (3/4\pi)\gamma^{\prime,\,2}(ZeB^\prime/mc)
\sim20(10^4{\rm \,s}/\delta t)^{1/7}$\,TeV.
The power radiated per particle is $\sim E/2\pi\Gamma^2 \delta t$.

Again considering BH-BH mergers and spherical shocks, 
the total received  synchrotron power 
 is  $L_{\rm synch}\sim\eta_{\rm CR}M_{\rm BH}c^2/(2\pi\delta t)
\sim3\times10^{51}(10^4{\rm s}/\delta t)(\eta_{\rm CR}/10^{-6})$\,ergs s$^{-1}$ for 
$M_{\rm BH}=10^8$\,M$_\odot$. 
If the particles have an $E^{-2}$ spectrum extending to $E_{\rm max}
(\delta t=10^4\,\,{\rm s})$, 
$\nu L_\nu^{\rm synch}\sim L_{\rm synch}/(2\ln(E_{\rm max}/mc^2))
\sim5\times10^{49}\,{\rm ergs \,\,s^{-1}}$
at $E_{\rm synch}$ with $\nu L_\nu\propto\nu^{1/2}$ for $E<E_{\rm synch}$.  
The observable rate of such events  
would be $\sim10$\,yr$^{-1}$, and the flux 
would be $\sim5\times10^{-8}$\,GeV s$^{-1}$ cm$^{-2}$ 
at a luminosity distance of $\sim7$\,Gpc.  Such
an event could have the appearance of a 
long-duration gamma-ray burst (GRB), and would
be observable by the {\it Fermi} and {\it Swift} satellites.
Whether or not a sub-class of long-duration GRBs might be 
ascribed to such a mechanism is unknown, but unlikely.  
Similar statements could be made for the analogous events
from lower mass BH-BH mergers,
and the association with short-duration GRBs.  More likely,
the lack of such events may limit $\eta_{\rm CR}$ to be
$\ll10^{-6}$, $\epsilon_{\rm B}$ to be $\ll1$, 
or $\Gamma\ne\Gamma_{\rm crit}$ 
in BH-BH mergers.

\textit{Conclusion.---}
We have derived robust, experimentally testable 
upper limits to the energies of primary UHECRs
accelerated by relativistic shocks driven by
astrophysical engines.  For such an engine to 
accelerate particles to $E_{\rm max}\sim10^{24}$\,eV (Eq.~\ref{emax_hubble})
it must have $L=L_{\rm max}$, and the relativistic
outflow it drives must reach $\Gamma=\Gamma_{\rm crit}$,
with $\epsilon_{\rm B}=1$.  For highly 
beamed flows,  $E_{\rm max}\sim10^{27}$\,eV (Eq.~\ref{kardashev}),
with the additional requirement that the solid angle for the flow  
must be $\lesssim10^{-12}$\,sr, likely unattainable in real astrophysical engines.
If any of these conditions
is not met, $E_{\rm max}$ is not reached.  Particles
reaching $E_{\rm max}$ suffer photomeson
or photodisintegration losses, producing secondary
neutrinos with average energy $\sim20$ times lower
than $E_{\rm max}$ for protons, observable from across the 
universe.  Thus, if a single UHE neutrino 
is  detected above $E_{\rm max}$ it will strongly imply
the existence of an unknown class of loss-less maximal
electromagnetic accelerators \cite{kardashev}, or 
physics beyond the Standard Model. 

We thank E.\ Waxman, J.\ Beacom, M.\ Kistler, S.\ Horiuchi, 
D.~Zhang, C.~Rott, and K.\ Murase for stimulating conversations, and  
C.~Dermer, E.\ Grashorn, N.~Lehtinen, 
A.\ Connolly, and S.~Buitink for helpful correspondence.
TAT~gratefully acknowledges 
A.\ Socrates for teaching him 
Eq.\ (\ref{lmax}). TAT is supported in part by an 
Alfred P.~Sloan Foundation Fellowship. BCL is supported by an
Elizabeth Clay Howald Presidential Fellowship.

\end{document}